\documentclass[12pt,preprint]{aastex}

\shorttitle{H-band Drop-out Galaxies}
\shortauthors{Huang et al.}

\begin{document}


\title{Four IRAC Sources with an Extremely Red H$-$[3.6] Color: Passive or Dusty Galaxies at \lowercase{z}$>$4.5? }


\author{J.-S. Huang\altaffilmark{1}, X. Z. Zheng\altaffilmark{2}, D. Rigopoulou\altaffilmark{3}, and G. Magdis\altaffilmark{3},G. G. Fazio\altaffilmark{1}, T. Wang\altaffilmark{1},\altaffilmark{4}}
%\affil{Harvard}

%\author{C. D. Biemesderfer\altaffilmark{4,5}}
%\affil{National Optical Astronomy Observatories, Tucson, AZ 85719}
%\email{aastex-help@aas.org}

%\and

%\author{R. J. Hanisch\altaffilmark{5}}
%\affil{Space Telescope Science Institute, Baltimore, MD 21218}


\altaffiltext{1}{Harvard-Smithsonian Center for Astrophysics, 60 Garden Str.,  Cambridge, MA02138, USA}
\altaffiltext{2}{Purple Mountain Observatory, 2 West Beijing Rd., Nanjing, Jiangsu Province, PRC}
\altaffiltext{3}{Department of Physics, Denys Wilkinson Building, Keble Road, Oxford, OX1 3RH, UK}
\altaffiltext{4}{Department of Astronomy, Nanjing University, Nanjing, Jiangsu Province, PRC}
%\altaffiltext{5}{Patron, Alonso's Bar and Grill}


\begin{abstract}
  We report detection of four IRAC sources in the GOODS-South field with an extremely red color of H$-$[3.6]$>$4.5. 
The four sources are not detected  in the deep HST WFC3 H-band image with H$_{limit}$=28.3\,mag. 
We find that only 3 types of SED templates can produce such a red H$-$[3.6] color: a very dusty SED with the Calzetti extinction of A$_V$=16 mag at z=0.8; a very dusty SED with the SMC extinction of A$_V$=8 mag at z=2.0$\sim$2.2;  and an 1Gyr SSP with A$_V \sim$0.8 at z=5.7. 
We argue that these sources are unlikely dusty galaxies at z$\leq$2.2 based on absent strong MIPS 24$\mu$m emission.
The old stellar population model at z$>$4.5 remains a possible solution for the 4 sources.  At z$>$4.5, these sources have stellar masses of
Log(M$_*$/M$_{\odot}$)=10.6$\sim$11.2. One source, ERS-1, is also a type-II X-ray QSO with  
L$_{2-8keV}$=1.6$\times$10$^{44}$ erg s$^{-1}$. 
One of the four sources is an X-ray QSO and another one is a HyperLIRG, suggesting a galaxy-merging scenario for the formation of these massive galaxies at high redshifts.

\end{abstract}


\section{Introduction}
  Extremely Red Objects(ERO) are of great interests to modern astrophysics. A simple red color criterion generally selects two types of galaxies: those at high redshifts and those with heavy dust extinction. With rapid progress in telescope apertures and detectors, this red color selection always leads to new types of galaxies or galaxies at record-high redshifts. After large format near-infrared array cameras became available for astronomical surveys, people started to detect galaxies with very red R$-$K colors \citep[EROs, R$-$K$>$5 or I$-$K$>$4]{elston1988}. EROs were so rare in the early days that they were thought to be abnormal objects at very high redshifts. There have been more and more EROs detected by larger aperture telescopes with more advanced IR array cameras. Most EROs with R$-$K$>$5 are now identified as elliptical and dusty galaxies at 
$0.6<z<1.5$ \citep{thompson1999,McCarthy2001, cimatti2002}. One extreme case, an ERO with I$-$K=6.5, was spectroscopically identified as a dusty Ultra-Luminous InfraRed Galaxy (ULIRG) at z=1.44 \citep{elbaz2002}. This source is analog to a local
ULIRG, Arp220. \citet{smail2002} suggested that most dusty EROs at high redshifts are LIRG/ULIRGs. The Spitzer IRAC permits very fast imaging of sky in mid-infrared bands with great depth. \citet{wilson2004} found that 17\% of their IRAC 3.6$\mu$m selected sample are EROs at z$\geq$1.

 Red color criteria are practically diversified, and applied to almost all kinds of photometry in optical and IR bands. But the physics for this type of criteria are limited to following: (1) the Lyman break at 912\AA; (2) the Balmer and the accumulated absorption line breaks at 3648\AA~~and 4000\AA; or (3) dust extinction. Red color caused by the Lyman/Balmer break can be used to estimate redshifts. 
In most deep broad band imaging surveys, one could not tell if a red color is due to Lyman/Balmer Break or dust extinction \citep{steidel2003}. 
An additional color in longer wavelength bands is usually applied together with the red color criterion to select galaxies at high redshifts.  For example, U$-$g and g$-$R colors were used to select galaxies at z=3 where the Lyman break shifts between U and g bands, commonly known as  U drop-out for red U$-$g color\citep{steidel2003}. The drop-out technique was applied in much longer wavelength bands to select galaxies at z=6$\sim$9. 
%\citep{eyles2005, Oesch2010,bunker2010,mclure2010,bouwens2010}. 
\citet{franx2003} used the NIR color J$-$K$>$2.3 to select Distant Red Galaxies (DRGs) with the strong Balmer/4000\AA~ break shifting in between the J and K at z$\sim$2.
The NIR spectroscopy for DRGs by \citet{kriek2007} shows that about half of their sample are passive evolved galaxies at z$\sim$2, 
and the rest are dusty galaxies in a much wider redshift range. Several groups idenitfied 24$\mu$m luminous and optically faint or invisible sources with R$-$[24]$>$14.2 ($f_{24}/f_R>1000$) as very dusty ULIRGs at z$\sim$2. These sources are confirmed spectroscopically by Spitzer IRS and ground-based optical spectroscopy \citep{houck2005, yan2007, dey2008, huang2009}.
 
  In this paper we report detection of four galaxies with extremely red colors of H$-$[3.6]$>$4.5 in the GOODS-South field.  Only one similar source,
a submillimeter galaxy (SMG) called GOODS 850-5 (aka GN10) in the GOODS-North field, was ever found to have H$-$[3.6]$>$4.5.  
This SMG was also detected by the Submillimeter Array \citep{wang2007} with a high angular resolution
of $\sim$2", permitting identification of its counterparts in shorter wavelength bands.
\citet{wang2009} performed ultra-deep J and H band imaging for this source with NIC3 camera on HST.  A total of 16 orbits of HST observation, reaching a nanoJansky depth in the F160W band, yields no detection for this source. Based on this red H$-$[3.6] color, they argue that its redshift is at z=4$\sim$6.5. Later, detection of CO(4-3) from this source confirms its redshift at z=4.05 \citep{daddi2009}. This study provides the first look at properties of this new type of object. 
More sources of this kind will be detected in the Cosmic Asembly Near-infrared Deep
Extragalactic Legacy Survey \citep[CANDELS]{grogin2011}.  

\section{Deep IR Imaging of GOODS-South}

\begin{figure}
\includegraphics[scale=.85]{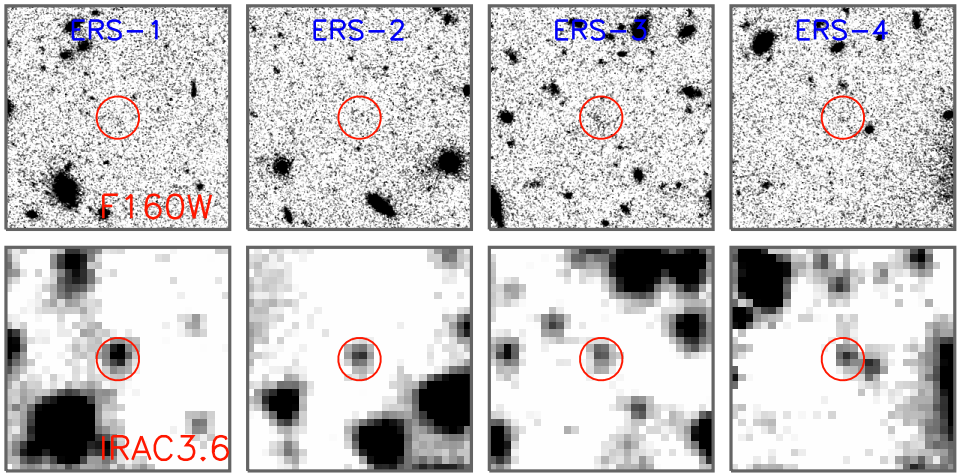}
\caption{Stamp images for the 4 H-band drop-out objects in the GOODS-South field in the F160W and 3.6$\mu$m bands. ERS-3 is marginally detected at 24, 250, 350$\mu$m, and 20cm. ERS-1 is an X-ray source detected in the Chandra 2Ms imaging \citep{alexander2003}.\label{f:stamp}}
\end{figure}

  The Great Observatories Origins Deep Survey(GOODS) is the deepest multi-wavelength survey with space telescopes including HST, Spitzer and Chandra\citep{goods}. The depth of GOODS IRAC 3.6$\mu$m
imaging reaches sub-microJansky level.  The deep NIR imaging of the 
GOODS-South field was selected for the Early Released Science \citep[ERS]{ers2010} for the Wide Field Camera 3 (WFC3), a fourth-generation UVIS/IR imager aboard HST.  We construct an H-selected multi-wavelength catalog including YJH+IRAC photometry 
in the ERS covered region.
The IR images have very different angular resolutions: 0.03" for the HST WFC3 YJH band images and $\sim$2" for the Spitzer IRAC 3.6-8.0 $\mu$m images.  A photometry program called TFIT is specifically designed to perform photometry on a lower resolution image with input information of 
object positions and light distributions measured in a high resolution image \citep{tfit2007}. 
The TFIT program convolves a PSF kernel to the high angular resolution stamp image for each object to construct lower angular resolution image templates and  fit them to the lower angular resolution image. 
In our case, we ran the TFIT to perform photometry on the IRAC 3.6$\mu$m image for the H-band selected galaxies detected in the ERS F160W image.  The TFIT also produces a residual image after subtracting all H-band detected galaxies in the 3.6$\mu$m image. We visually inspected the residual image and found four IRAC sources detected at 3.6$\mu$m with no H-band counterparts shown in Figure~\ref{f:stamp}. The limiting magnitude for the input
H-selected sample is H$=$28.3 mag at 3$\sigma$ level, therefore these sources are fainter than H$=$28.3 mag and have colors redder than H$-$[3.6]=4.5.

    We searched for counterparts of these sources in all available wavelength bands in the GOODS-South field. All four sources are detected in the remaining 3 IRAC bands.  None of these sources is detected in the GOOD-South ACS BVIZ images with the 5$\sigma$ limiting magnitudes of 28.65,28.76,28.17.and 27.93 respectively. The K-band is the only band available
in between H and 3.6$\mu$m, permitting further constraint on its SED and photometric redshift.   The 5$\sigma$ limiting magnitude for
the K-band image of the GOOD-South field \citep{esoKband} is 24.4 mag and none of our sources is detected in K band.
FIR observation is also critical in determining properties of these red sources \citep{wang2009}.   ERS-3 is clearly detected at 24$\mu$m and ERS-2 is marginally
detected at $\sim$3$\sigma$ level. We inspected the Herschel SPIRE deep imaging of the CDFS, and found only ERS-3 is 
marginally detected at 250$\mu$m and 350$\mu$m. The PSF for the SPIRE 500 $\mu$m band image is 
too broad (~36.6") to permit accurate extraction of flux density for  ERS-3 \citep{huang2011}. ERS-3 is also detected at 1.4gHz
with f$_{1.4gHz}$=29.2$\pm$8$\mu$Jy. The remaining three sources are not detected in radio with f$_{1.4gHz}<$24$\mu$Jy.
No submillimeter/millimemter source is detected in the locations of these four sources \citep{scott2009, weis2009}. 
Another source, ERS-1, is  an X-ray source in the Chandra 2Ms catalog\citep{alexander2003}. ERS-2 and ERS-4 are detected only in 4 IRAC bands (Table.1).

\section{SEDs, Photometric Redshifts, and Properties of the Extremely Red Objects}

  For three out of the four sources in this study, only NIR+IRAC flux densities are available for their photometric redshift estimation. 
The most predominant feature in their SEDs is the extremely red color of H$-$[3.6]$>$4.5. We first rule out that those sources are brown dwarves.  A brown dwarf with T=600K has  H$-$[4.5]$>$4.0 \citep{legget2010}, such a brown dwarf 
should have [3.6]$-$[4.5]$>$1 due to the methane absorption at 3.6$\mu$m in its photosphere. All our sources have [3.6]$-$[4.5]$<$0.5.
It is unlikely that this red color is due to the Lyman break at z$>$15. 
Galaxies with a strong Balmer/4000\AA~ break at 3$<$z$<$8 can have very red H$-$[3.6] colors. Recently \citet{johan2011} detected a lensed source at z=6.02 behind cluster A383 with H$-$[3.6]=1.5, arguing that this is a possible passive galaxy. A few more lensed galaxies with extreme red optical-MIR color
are identified to be dusty galaxies at either z$\sim$2 or z$>$6 \citep{boone2011,laporte2011}. Their H$-$[3.6] colors are only in range of 0.5$<$H$-$[3.6]$<$2.

\subsection{SED Fitting}

  We model the H$-$[3.6] color using stellar population models of both BC03 \citep[BC03]{bc03} and the upgraded model CB07 \citep{cb07} 
emphasizing on AGB star contribution in the rest-frame NIR bands. The templates are constructed with various stellar populations of the solar metallicity and a very wide range of dust extinction of 0$<$A$_V<$25.
The star formation history used in constricting the  template set includes single burst, exponential decreasing with various e-fold times, and constant rates.
Two types of extinction curves are used in the SED templates: the widely used Calzetti extinction curve for galaxies \citep{calzetti2000} and the SMC extinction curves \citep{gordon2003}. Both extinction curves are only up to 2.2$\mu$m, while our detected photometry points for the four sources are in 3.6$<\lambda<$8.0$\mu$m. We extend both curve to the IRAC bands using the MIR dust extinction curve in 3.6$<\lambda<$24$\mu$m \citep{chapman2009}.

  We first fitted our model templates to GOODS 850-5 to investigate what kind of stellar population and how much amount of dust extinction can
make such a red H$-$[3.6] color. GOODS 850-5 is already known at z=4.05, thus provides better constraint on stellar population and dust extinction. Both BC03 and CB07 models with either Calzetti or SMC extinction yields a similar result for GOODS 850-5 : 1Gyr old single stellar population 
with modest extinction of A$_V$=2.4$\sim$3.6.  The best fit template is an 1Gyr single stellar population model with the Calzetti extinction of A$_V$=3.6. \citet{wang2009} obtained a similar model template for their best fitting but yielding much higher redshifts at z=6.9.

\begin{figure}
\plotone{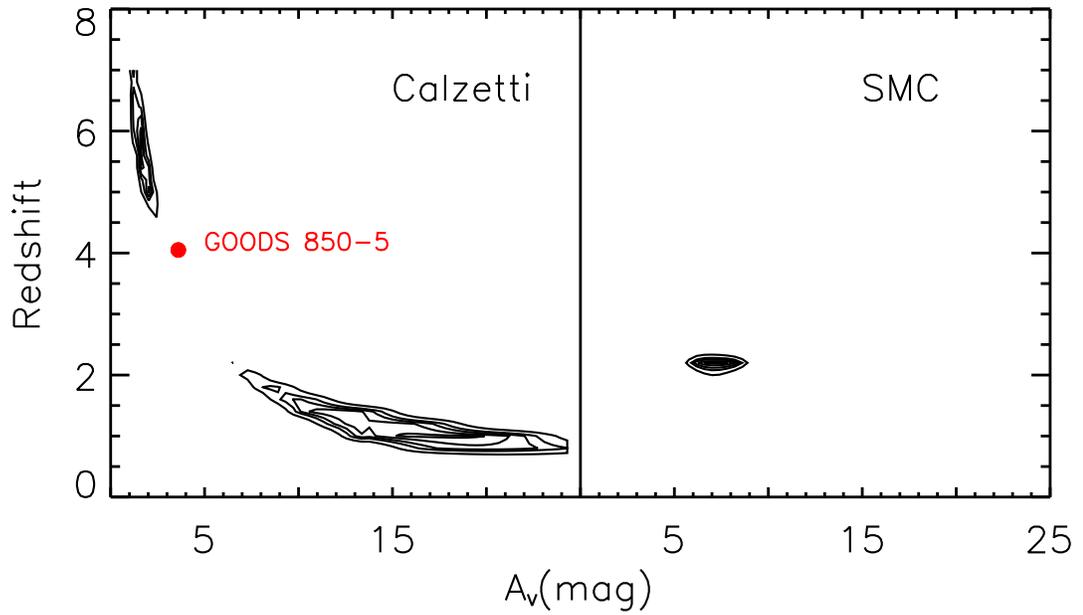}
\caption{The likelihood contours as a function of redshift and dust extinction A$_V$. for the best-fit SED of ERS-1. SED fitting for the 
remaining objects yields the same solutions.
The left panel is the contour  with the 1Gyr stellar population model and the Calzetti extinction and the right panel with the SMC extinction. 
The 1Gyr stellar population model with the Calzetti extinction of A$_V$=3.6 is also the best fit for GOODS 850-5 at z=4.05.
\label{f:kai_con}}
\end{figure}

 We argue that the four objects in this study  are at the same redshifts: they have very similar SEDs, and their positions are very close to each other, with a mean distance of $\sim$1.5' to their closest neighbors. We fitted the SED templates to the six IR photometry points (H, K, and 4 IRAC bands) for each object in the sample.  
Our fitting yields two extreme solutions with the Calzetti exinction: a very dusty
template with A$_V$=16$\sim$18 at z$\sim$0.8 and an old stellar population template with z$\sim$5.7 and A$_V \sim$0.8 (Figure~\ref{f:kai_con}).  
The templates with the SMC extinction yields a similar dusty solution with A$_V$=7$\sim$8 at z$\sim$2.2.    
By applying heavy dusty extinction with A$_V>$7 to templates, its shape  and the resulting photometric redshift are only 
determined by extinction curves. 
For example, the SMC extinction curve yields a photometric redshift of z$_p$=2.2 for our objects, which is caused by
a dip at 1.25$\mu$m in the SMC extinction curve(A(1.65$\mu$m)/A$_V$=0.169, A(1.25$\mu$m)/A$_V$=0.131, and A(0.81$\mu$m)/A$_V$=0.567, \citet{gordon2003}). With a very high A$_V$ value, this feature is amplified. At z=2.2, this extinction dip shifts to the IRAC 3.6$\mu$m band to 
make H$-$[3.6] redder and [3.6]$-$[8.0] bluer.  The photometric redshifts obtained with the SMC extinction curve are mainly driven by this feature.  

 There are generally three final solutions (Figure~\ref{f:kai_con}) in our SED fitting for the 4 objects:  a dusty template at z=0.8 with 
the Calzetti extinction of A$_V$=16;  a dusty template at z=2.2 with the SMC extinction of A$_V$=8;  and an old stellar population template 
with age of 1Gyr and A$_V <$1. Though each solution has a slightly different minimum $\chi$ of 3$<\chi_{min}<$6, we consider each solution equally possible. All three SED models are able to produce H$-$[3.6]$>$4.5, 
and require extreme conditions in the galaxies: either extremely dusty of A$_V>$7 or even A$_V$=16 or very massive galaxies at z$>$4.5,
both of which are very rare in current extragalactic surveys.
 
\subsection{Extremely Dusty Galaxies at z$<$3?}
  
 In the first solution, a galaxy with the Calzetti extinction of A$_V \sim$16 at z=0.8 can have H$-$[3.6]$>$4.5. At z=0.8, the IRAC 3.6$\mu$m band probes the rest-frame K-band.
Our sources have 3.6$\mu$m flux density of f$_{3.6}$=0.6$\sim$1.5$\mu$Jy, implying that their stellar masses are less than 
5$\times$10$^9$ M$_{\odot}$. Most galaxies with such a small stellar mass at z=0.8 are
blue galaxies with no dust extinction. M82 is a dusty galaxy with lower stellar mass of
4$\times$10$^9$ M$_{\odot}$, with heavy dust
obscuration (5$<$A$_V<$51) only occurring in its center \citep{beirao2008}. The whole M82 appears much bluer than our objects. 
The H$-$[3.6] color at z=0.8 is equivalent to the rest-frame I$-$K color. M82 has I$-$K=0.82 \citep{dale2007}, because
most stars in the disk of M82 are  in the outside of its dusty region.
Thus an object like M82 at z=0.8 would be detected in the ERS H-band imaging.  
This solution, however, requires that the whole galaxy should be in heavy obscuration. Only ULIRGs have such a obscured morphology. 
Using M82 central region SEDs,
we predict that the MIPS 24$\mu$m flux densities for the first 3 sources would have f$_{24}$=50$\sim$70$\mu$Jy, well above the
FIDEL MIPS 24$\mu$m limiting flux density. Only ERS3 is marginally detected at 24$\mu$m with f$_{24}$=39$\mu$Jy, the rest are not detected at 
24$\mu$m. We argue that this scenario is least possible.

  The second solution is the template with the SMC extinction of A$_V$=7$\sim$8 at z=2$\sim$2.2. There are many Dust Obscured Galaxies(DOG) identified at z$\sim$2 \citep{houck2005,dey2008,bussmann2009}. Several groups \citep{houck2005, yan2007, huang2009} performed mid-infrared spectroscopy 
for DOGs and detected a very strong silicate absorption feature at 9.8$\mu$m in their spectra, indicating a very heavy dust extinction.
\citet{bussmann2009} took high resolution H-band imaging of these sources in the Bootes field with NICMOS on HST to study their rest-frame optical morphologies. We identified these DOGs in the Bootes IRAC photometry catalog \citep{ashby2009}, and found that DOGs were
generally red with 1.5$<$H$-$[3.6]$<$3.3,
and luminous at 3.6$\mu$m with f$_{3.6}$=5$\sim$60$\mu$Jy.
The four sources in this study are much fainter at 3.6$\mu$m, and have a much redder color of H$-$[3.6]$>$4.5. 
On the other hand, the DOGs in \citet{bussmann2009} have a 24-to-8$\mu$m flux ratio of 
f$_{24}$/f$_{8}$=40$\sim$350. If the 4 sources were indeed fainter DOGs at z$\sim$2, they should have a 24$\mu$m flux density 
of f$_{24} >$120$\mu$Jy, much higher than the 24$\mu$m limiting flux density in GOODS-South. Based on much redder H$-$[3.6] and fainter 
24$\mu$m flux,  we argue that these objects are at higher redshifts than the DOGs at z$\sim$2.

\subsection{Massive  Galaxies at z$>$4.5?}

\begin{figure}
\epsscale{0.70}
\plotone{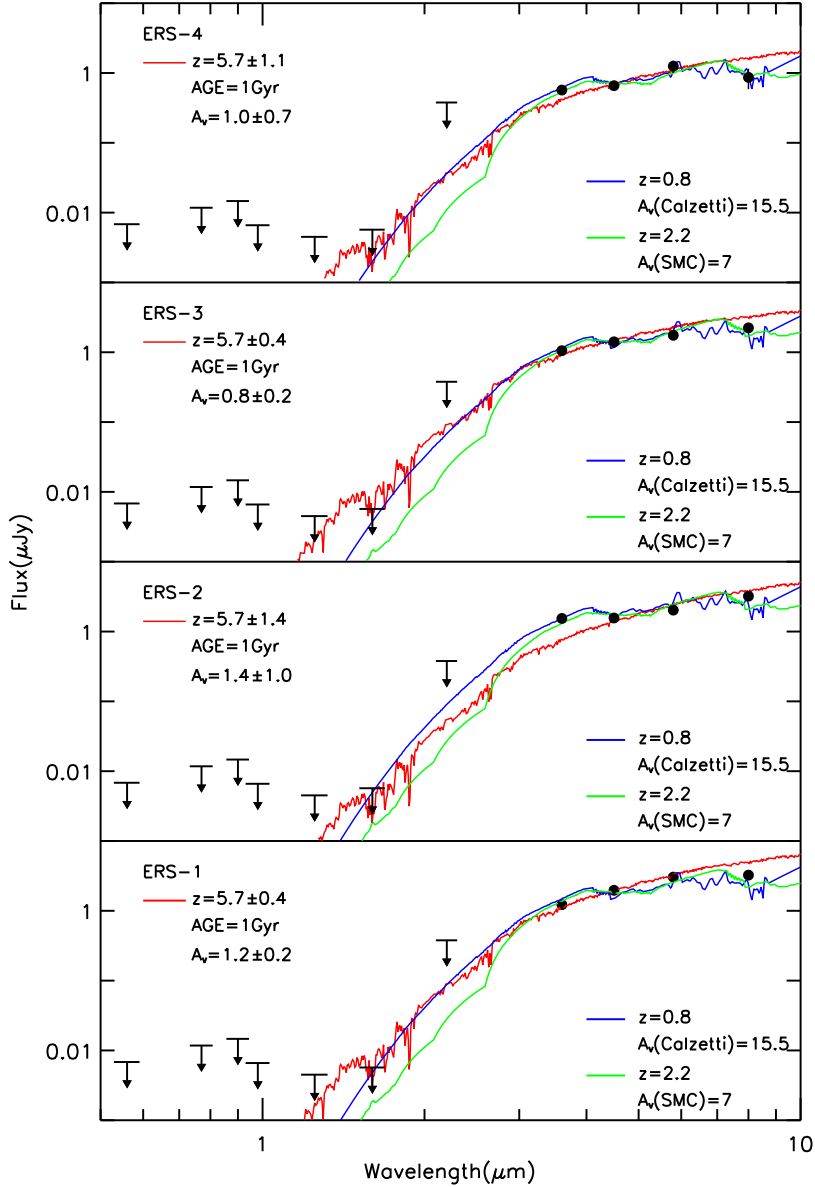}
\caption{Optical-to-MIR SEDs for the H-band drop-out sources in ERS. The best-fit templates to the SEDs are dusty SSP models with
E(B-V)=0.2$\sim$0.35 and $z_p=5.7$. The fitting also yields stellar mass of Log(M$_*$/M$_{\odot}$)=10.6$\sim$11.2 for the 4 sources.
We also plot two dusty model templates against observed SEDs.
\label{f:sed}}
\end{figure}

 The 1Gyr SSP template with the Calzetti extinction of A$_V$=0.8 at z$\sim$5.7 can also fit the SEDs of the four objects, very similar to 
the best-fit template for GOODS 850-5. In this scenario, the red H$-$[3.6] is mainly due to the Balmer/4000\AA~~jump shifting in between
H and 3.6$\mu$m bands at z$>$4.  Figure~\ref{f:sed} shows
that the old stellar template fits to SEDs of the 4 objects. The resulting photometric redshifts have a large error of 0.4$<\sigma$(z)$<$1.1. We argue
that the 4 objects are at the same redshifts. Thus by adopting the best $\sigma (z)$ in the 4 objects, these objects should  be at  z$>$4.5 at 
3$\sigma$ level.

ERS-3 is detected at 250 and 350$\mu$m, thus has a very strong FIR emission, very similar to GOODS 850-5 \citep{huang2011}.
The best fit SED models for both GOODS 850-5 and ERS-3 are old stellar population models.
We propose a two-component SED model to reconcile the old stellar population and FIR emission in these objects: an old stellar population 
and a very dusty star-forming component. The star-forming component is so dusty, simliar to those dusty galaxies detected at z$\sim$2 
\citep{houck2005,yan2007,dey2008, huang2009}, that its optical/NIR SED is dominated by the old stellar population component.
For example, a dusty component with the same stellar mass and A$_V$=6  at z$>$5 would only contribute 10\% increase at 8 micron, 
and much lower percentage in the shorter IRAC bands. 
Assuming a typical dust temperature of T$_{dust}$=40K and redshift of z=5.7, we calculate the FIR luminosity for ERS-3 
as Log(L$_{FIR}$/L$_{\odot})$=13.1. The FIR-to-Radio flux ratio is about q=2.26, consistent with q values for 
submillimeter galaxies at $z>4$ \citep{huang2011}. The remaining 3 objects are not be detected by Herschel SPIRE, but may still be IR luminous galaxies with just Log(L$_{FIR}$/L$_{\odot})<$13.1.

  The SSP model fitting also yields stellar mass of Log(M$_*$/M$_{\odot}$)=10.6$\sim$11.2 for our sources 
(Figure~\ref{f:sed}). Spectroscopic confirmed galaxies at z$\sim$5.7 including both Lyman-break (LBGs) and Ly-$\alpha$ (LAE) galaxies have a typical stellar mass of Log(M$_*$/M$_{\odot}$)$\leq$10 \citep{yan2006,lai2007, young2007, johan2011}. Recently \citet{marchesini2010} argued that very massive galaxies were already formed at 3$<$z$<$4. Theoretically, \citet{li2007} argued that QSOs at z$>$6 resided in a massive halo of 
M$\sim$8$\times$10$^{12}$M$_{\odot}$, and stellar mass for their host galaxies can be as high as 10$^{12}$M$_{\odot}$.
We have another piece of evidence consistent with these systems being massive.  ERS-1 is also an X-ray source detected in the Chandra 2Ms survey \citep{alexander2003}. It has x-ray flux densities of $f_{0.5-2keV}=7.55\pm2.14\times 10^{-17} erg~s^{-1}~cm^{-2}$ and $f_{2-8keV}=6.50\pm1.50\times 10^{-16} erg~s^{-1}~cm^{-2}$. The hard X-ray Luminosity for this source is L$_{2-8keV}$=1.6$\times$10$^{44} erg~s^{-1}$ at $z_p=5.7$  assuming no absorption correction.
Its X-ray-to-optical-flux ratio ($f_x/f_R$) is higher than 60 and the hardness ratio is $\sim$1, thus ERS-1 is an obscured type-II QSO. A typical black hole mass for such a QSO at z=3$\sim$6 is 10$^9$ $\sim$ 10$^{10}$ M$_{\odot}$ \citep{netzer2003,shemmer2004,fan2006}. Assuming a typical H${\alpha}$ FWHM of 2000 km/s for a QSO, we convert the L$_{2-8keV}$  to black hole mass for ERS-1 as M$_{BH}$=5$\times$10$^{8}$ M$_{\odot}$ using the relation proposed by \citet{sarria2010}. \citet{trakhtenbrot2010} argued that a host galaxy with 10$^9$ M$_{\odot}$ black hole has a typical stellar mass of M$_*\sim$10$^{11}$ M$_{\odot}$, consistent with the stellar mass we derived for ERS-1. 

\section{Summary and Discussion}

  We identified four IRAC sources in the GOODS-South field with extremely red color of H$-$[3.6]$>$4.5. 
The only known source with a similar H$-$[3.6] color is GOODS 850-5, a SMG in the GOODS-North field. 
We argue that the four sources must be at the same redshift based on the following facts: they have similar rest-frame optical/NIR SEDs; and they are spatially very close to each other with a mean angular distance $\sim 1.5'$.
Only 3 types of templates can produce H$-$[3.6]$>$4.5: a very dusty template with the Calzetti extinction of A$_V$=16 mag at z=0.8; a very dusty templates with the SMC extinction of A$_V$=8 mag at z=2.0;  and an 1Gyr SSP model with A$_V \sim$0.8 at z=5.7. 
By comparing the 4 objects with local dusty galaxies and DOGs at z$\sim$2, we argue that they are unlikely dusty galaxies at z=0.8 or z=2.2 based on absent strong 24$\mu$m emission.
The old stellar population model at z$>$4.5, with the best fit at z=5.7, remain a possible solution for the 4 sources. One of our sources, ERS-3, is also detected by Herschel at 250$\mu$m and 350$\mu$m, yielding Log(L$_{FIR}$/L$_{\odot}$)=13.2.  
We propose a two-component SED model for these sources: an old SSP component dominating their optical-to-MIR 
SEDs and a very dusty star-forming component mainly contributing to their FIR SEDs. The SED fitting yields stellar masses of
Log(M$_*$/M$_{\odot}$)=10.6$\sim$11.2 for the four sources. One source, ERS-1, is also a type-II x-ray QSO with  
L$_{2-8keV}$=1.6$\times$10$^{44} erg~s^{-1}$. Based on the M$_{BH}$-M$_{bulge}$ relation for high-z QSOs, ERS-1 should have a massive bulge 
of Log(M$_*$/M$_{\odot}$)=11. One of the four sources is an X-ray QSO and another one is a HyperLIRG, 
suggesting a galaxy-merging scenario for the formation of these massive galaxies at high redshifts.

\acknowledgments

This work is based on observations made with the {\it Spitzer Space Telescope}, which is operated by the Jet Propulsion Laboratory, California Institute of Technology under NASA contract 1407,  and with the NASA/ESA HST obtained at the Space Telescope Science Institute, which is operated by the association of Universities for Research in Astronomy, Inc., under NASA contract NAS 5-26555. 

%We are grateful to V. Barger, T. Han, and R. J. N. Phillips for
%doing the math in section~\ref{bozomath}.
%More information on the AASTeX macros package is available \\ at
%\url{http://www.aas.org/publications/aastex}.
%For technical support, please write to
%\email{aastex-help@aas.org}.


%% copy editing.  Individual instruments can be provided in parentheses,
%% after the keyword, but they will not be verified.

Facilities: \facility{Spitzer(IRAC)}, \facility{HST(STIS)}, \facility{CXO(ASIS)}.

\clearpage

%% the file, uncomment the \plotone, \plottwo, and
%% If you need a layout that cannot be achieved with \plotone or
%% \plottwo, you can invoke the graphicx package directly with the
%% \includegraphics command or use \plotfiddle. For more information,
%% please see the tutorial on "Using Electronic Art with AASTeX" in the
%% documentation section at the AASTeX Web site,
%% http://www.journals.uchicago.edu/AAS/AASTeX.
%%
%% The examples below also include sample markup for submission of
%% supplemental electronic materials. As always, be sure to check
%% the instructions to authors for the journal you are submitting to
%% for specific submissions guidelines as they vary from

%% This example uses \plotone to include an EPS file scaled to
%% 80% of its natural size with \epsscale. Its caption
%% has been written to indicate that additional figure parts will be
%% available in the electronic journal.

%\begin{figure}
%\epsscale{0.70}
%\plotone{fig1.ps}[angle=90]
%\caption{Stamp images for the 4 objects in the GOODS-South field in ACS F774W, WFC3 F160W, IRAC 3.6-8.0$\mu$m,
%MIPS 24$\mu$m bands. ERS-3 is marginal detected at 24$\mu$m, and this source is detected by Herschel at 250 and 350$\mu$m.
%ERS-1 is an X-ray source detected in Chandra 2Ms survey\citep{alexander2003}.\label{f:stamp}}
%\end{figure}

\clearpage

%% Here we use \plottwo to present two versions of the same figure,
%% one in black and white for print the other in RGB color
%% for online presentation. Note that the caption indicates
%% that a color version of the figure will be available online.
%%

%% This figure uses \includegraphics to scale and rotate the still frame

%\begin{figure}
%%\includegraphics[angle=90,scale=.50]{f3.eps}
%\caption{Animation still frame taken from \citet{kim03}.
%This figure is also available as an mpeg
%animation in the electronic edition of the
%{\it Astrophysical Journal}.}
%\end{figure}

%% If you are not including electonic art with your submission, you may
%% mark up your captions using the \figcaption command. See the
%% No more than seven \figcaption commands are allowed per page,
%% so if you have more than seven captions, insert a \clearpage

%% Tables should be submitted one per page, so put a \clearpage before

%% Two options are available to the author for producing tables:  the
%% deluxetable environment provided by the AASTeX package or the LaTeX
%% table environment.  Use of deluxetable is preferred.
%%

%% Three table samples follow, two marked up in the deluxetable environment,
%% one marked up as a LaTeX table.

%% In this first example, note that the \tabletypesize{}
%% command has been used to reduce the font size of the table.
%% We also use the \rotate command to rotate the table to
%% landscape orientation since it is very wide even at the
%% Note also that the \label command needs to be placed

%% This table also includes a table comment indicating that the full
%% version will be available in machine-readable format in the electronic
\clearpage

%\begin{turnpage}
\begin{deluxetable}{cccccccccr}
\tabletypesize{\scriptsize}
\tablecaption{Infrared flux densities for the H-band Dropout sources\label{tbl-1}}
\tablewidth{0pt}
\tablehead{
\colhead{Name} & \colhead{RA} & \colhead{DEC} & \colhead{F160W} & \colhead{K} & \colhead{3.6$\mu$m} & \colhead{4.5$\mu$m} &
\colhead{5.8$\mu$m} & \colhead{8.0$\mu$m} & \colhead{24$\mu$m}
}
\startdata
ERS-1 & 53.084726 & -27.707964 & 0.0066$\pm$0.002 & 0.2904$\pm$0.1398 &1.23$\pm$0.03 & 1.97$\pm$0.04 & 3.04$\pm$0.24 & 3.25$\pm$0.26 & -11.7$\pm$3.8\\
ERS-2 & 53.132749 & -27.720144 & 0.0036$\pm$0.002 & -0.0431$\pm$0.1681&1.54$\pm$0.20 & 1.56$\pm$0.10 & 2.03$\pm$0.26 & 3.21$\pm$0.28 &  14.7$\pm$4.0\\
ERS-3 & 53.060827 & -27.718263 & 0.0019$\pm$0.002 & 0.0727$\pm$0.1398 &1.05$\pm$0.03 & 1.41$\pm$0.05 & 1.75$\pm$0.23 & 2.24$\pm$0.25 &  39.5$\pm$4.3\\
ERS-4 & 53.167161 & -27.715316 & 0.0054$\pm$0.002 & 0.2628$\pm$0.1382&0.57$\pm$0.06 & 0.66$\pm$0.06 & 1.25$\pm$0.25 & 0.87$\pm$0.28 &   5.9$\pm$3.6\\
\enddata

\tablecomments{All flux densities in this table are in unit of $\mu$Jy.}
\end{deluxetable}
%\end{turnpage}

%% Text for table notes should follow after the \enddata but before
%% the \end{deluxetable}. Make sure there is at least one \tablenotemark
%% in the table for each \tablenotetext.

\end{document}